\documentstyle[12pt]{article}
\font\twelveof=msym10 at 12pt

\def\case#1#2{{\textstyle{#1\over #2}}}
\def\sech{\mathop{\rm sech}\nolimits}
\def\cosech{\mathop{\rm cosech}\nolimits}
\def\N{\mbox{\twelveof N}}
\newcommand{\trait}{\rule{0.5cm}{0.1mm}}

\title{Zero-energy states for a class of quasi-exactly solvable rational potentials}
\author{B. Bagchi $^{a,}$\thanks{E-mail: bbagchi@cubmb.ernet.in}\ , C. Quesne
$^{b,}$\thanks{Directeur de recherches FNRS; E-mail: cquesne@ulb.ac.be} \\
{\small \sl $^a$ Department of Applied Mathematics, University of Calcutta,} \\
{\small \sl 92 Acharya Prafulla Chandra Road, Calcutta 700 009, India}\\ 
{\small \sl $^b$ Physique Nucl\'eaire Th\'eorique et Physique Math\'ematique, 
Universit\'e Libre de Bruxelles,} \\ {\small \sl Campus de la Plaine CP229,
Boulevard~du Triomphe, B-1050 Brussels, Belgium}}
\date{ }
\begin{document}
\baselineskip=22pt plus 1pt minus 1pt
\maketitle

\begin{abstract}
Quasi-exactly solvable rational potentials with known zero-energy solutions of the
Schro\" odinger equation are constructed by starting from exactly solvable
potentials for which the Schr\" odinger equation admits an so(2,1) potential algebra.
For some of them, the zero-energy wave function is shown to be normalizable and to
describe a bound state.
\end{abstract}

\vspace{0.5cm}

\hspace*{0.3cm}

\hspace*{0.3cm}
\newpage
%
%
In quantum mechanics the number of exactly solvable~(ES) potentials being
limited, there have been recent searches for either
quasi-exactly solvable(QES)~\cite{turbiner} or conditionally exactly
solvable (CES)~\cite{dutra93, dutra94, grosche,dutt} systems, for which the
energy spectrum is partly or completely known under certain constraint conditions
among the potential parameters, respectively. Apart from being mathematically
interesting, both types of potentials offer useful insights into the description of
physical phenomena: for instance, one can establish a correspondence between
QES~problems and the spin-boson and spin-spin interacting
models~\cite{zaslavskii}, while CES~potentials may be shown to be related to the
ES Coulomb, anharmonic, and Manning-Rosen potentials~\cite{dutt}.\par
%
%
There has recently been a resurgence of interest~\cite{daboul} in tracking down
potentials with zero binding energy. For one thing, an $E=0$ quantal state is a
QES~system in its own right; for another, familiarity with the Coulomb problem
persuades one to expect that all zero-energy states lie in the continuum and
so are not normalizable. However, Daboul and Nieto~\cite{daboul} have made a
systematic survey with radial power law potentials to arrive at several
exceptional solutions, which include two cases where the $E=0$ wave function is
normalizable and the corresponding state is either bound or unbound. Bound
$E=0$ solutions have also been reported before. For example, within the framework
of the general effective radial potentials for an interacting spin 1/2 particle,
Barut~\cite{barut} had pointed out long ago that exact zero-energy solutions could
exist with properties of confinement (normalizable case) and leakage
(non-normalizable case).\par
%
%
The purpose of this letter is to present another evidence of zero-energy
normalizable solutions for a class of QES rational potentials. Our approach for
constructing such a class combines elements of various methods for generating ES
or CES~potentials, including algebraic techniques based upon the use of so(2,1) as a
potential algebra for the Schr\"odinger equation~\cite{miller,wu,cq}.\par
%
%
A few remarks on so(2,1) are in order. As shown by Wu and Alhassid~\cite{wu}, the
underlying commutation relations of so(2,1), namely $[J_+, J_-] = -2J_0$, $[J_0,
J_{\pm}] = \pm J_{\pm}$, may be given a differential realization $J_0 = - i
\partial/\partial\phi$, and $J_{\pm} = e^{\pm i \phi} \left[\pm (\partial/
\partial x) + F(x) \left(i \partial/\partial \phi \mp \frac{1}{2}\right) + G(x)
\right]$, where the two functions~$F$ and~$G$ satisfy coupled differential
equations
\begin{equation}
  F' = 1 - F^2, \qquad \mbox{and} \qquad G' = - F G,     \label{eq:so(2,1)-cond}
\end{equation}
dashes denoting derivatives with respect to $x$. The Casimir operator $J^2 = J_0^2
\mp J_0 - J_{\pm} J_{\mp}$ being explicitly $J^2 = \left(\partial^2/ \partial x^2
\right) - F' \left[\left(\partial^2/ \partial \phi^2\right) + \frac{1}{4}
\right] + 2 i G' \left(\partial/ \partial \phi\right) - G^2 - \frac{1}{4}$, it
follows readily that an irreducible representation of the potential algebra so(2,1)
has basis states, which are eigenfunctions of different Hamiltonians, but conform
to the same energy level. In other words, with $J_0 |km\rangle = m |km\rangle$,
and $J^2 |km\rangle = k(k-1) |km\rangle$ ($m=k$, $k+1$, $k+2$,~$\ldots$), where
$|km\rangle = \psi_{km}(x) e^{im\phi}$ are the basis functions, the
coefficient functions $\psi_{km}(x)$ obey the Schr\"odinger equation
\begin{equation}
  - \psi_{km}'' + V_m \psi_{km} = - \left(k - \case{1}{2}\right)^2 \psi_{km},  
\end{equation}
where the one-parameter family of potentials is
\begin{equation}
  V_m = \left(\case{1}{4} - m^2\right) F' + 2m G' + G^2.   \label{eq:V_m}
\end{equation}
\par
%
%
Wu and Alhassid~\cite{wu} have considered particular solutions
of~(\ref{eq:so(2,1)-cond}) to deal with Morse, P\"oschl-Teller, and Rosen-Morse
potentials from~(\ref{eq:V_m}). On the other hand, Englefield and
Quesne~\cite{cq} have explored more general possibilities to identify three classes
of solutions from~(\ref{eq:so(2,1)-cond}) according to whether $F^2 < 1$, $F^2 = 1$
or $F^2 > 1$:
\begin{equation}
  \begin{array}{lll}
     \mbox{(I)} & F(x) = \tanh x, \quad & G(x) = b \sech x, \\[0.2cm]
     \mbox{(II)} & F(x) = \pm 1, \quad & G(x) = b e^{\mp x}, \\[0.2cm] 
     \mbox{(III)} & F(x) = \coth x, \quad & G(x) = b \cosech x.
  \end{array}   \label{eq:3cases}
\end{equation}
\par
%
%
Substitution of these solutions in~(\ref{eq:V_m}) leads to non-singular
Gendenshtein, Morse, and singular Gendenshtein potentials. These solutions
encompass those obtained by Wu and Alhassid~\cite{wu} in that Gendenshtein
potentials are disguised versions of P\"oschl-Teller potentials; further, for
particular values of the parameters, one gets non-singular Rosen-Morse potentials
from singular Gendenshtein potentials.\par
%
%
We now outline a procedure to obtain QES potentials from a very general scheme,
which addresses the problem of generating ES potentials in quantum
mechanics~\cite{sudarshan, levai}.\par
%
%
Let us consider a change of variables $x \to f(u)$ resulting in the Schr\"odinger
wave function transforming as $\psi(x) \to g(x) \chi(u(x))$. The standard form of the
Schr\"odinger equation
\begin{equation}
  - \frac{d^2 \psi}{dx^2} + [V(x) - E] \psi(x) = 0   \label{eq:schrodinger}
\end{equation}
is thus modified to a more general expression
\begin{equation}
  \frac{d^2 \chi}{du^2} + Q(u) \frac{d\chi}{du} + R(u) \chi(u) = 0.
  \label{eq:mod-schrodinger} 
\end{equation}
The functions $Q(u)$ and $R(u)$ are given by
\begin{eqnarray}
  Q(u) & = & \frac{u''}{u^{\prime 2}} + \frac{2g'}{gu'}, \label{eq:Q} \\
  R(u) & = & \frac{g''}{gu^{\prime 2}} + \frac{E - V(x)}{u^{\prime 2}}, \label{eq:R}
\end{eqnarray}
where the dashes denote derivatives with respect to $x$.\par
%
%
{}From~(\ref{eq:Q}) and~(\ref{eq:R}), the difference $E - V(x)$ may be represented in
terms of the function~$u$ only:
\begin{equation}
  E - V(x) = \frac{1}{2} \Delta V(u) + u^{\prime 2} \left(R - \frac{1}{2} \frac{dQ}{du}
  - \frac{1}{4} Q^2\right),   \label{eq:E-V}
\end{equation}
where the quantity $\Delta V(u)$ is the so-called Schwartzian
derivative~\cite{aly},
\begin{equation}
  \Delta V(u) = \frac{u'''}{u'} - \frac{3}{2} \left(\frac{u''}{u'}\right)^2.
  \label{eq:Delta}
\end{equation}
\par
%
%
It is clear from~(\ref{eq:E-V}) that if the functions $Q(u)$ and $R(u)$ are known
explicitly, we can try out various choices of~$u(x)$ to arrive at an ES~potential. In
the literature, several studies~\cite{sudarshan,levai} of Eq.~(\ref{eq:E-V}) have
been made by comparing Eq.~(\ref{eq:schrodinger}) with those differential
equations whose solutions are analytically known.\par
%
%
We would like to point out that CES or QES~potentials may also be obtained from the
secondary differential equation~(\ref{eq:mod-schrodinger}) by putting $Q(u) = 0$.
The latter implies $u' g^2 = \mbox{constant}$, $\psi = (u')^{-1/2} \chi(u(x))$, and
from~(\ref{eq:E-V}) or (\ref{eq:R})
\begin{equation}
  E - V(x) = \case{1}{2} \Delta V(u) + R u^{\prime 2}.   \label{eq:E-V-special}
\end{equation}
\par
%
%
To get a meaningful representation of~(\ref{eq:E-V-special}), we set $R = E_T -
V_T(u)$, and use~(\ref{eq:Delta}) along with the transformation $x = f(u)$ to get
the result
\begin{equation}
  E_T - V_T(u) = [E - V(f(u))] (f'(u))^2 + \case{1}{2} \Delta V(f(u)),
  \label{eq:E_T-V_T}
\end{equation}
where
\begin{equation}
  \Delta V(f(u)) = \frac{f'''(u)}{f'(u)} - \frac{3}{2} \left(\frac{f''(u)}{f'(u)}\right)^2.
  \label{eq:Delta-spec}
\end{equation}
In Eqs.~(\ref{eq:E_T-V_T}) and~(\ref{eq:Delta-spec}), the dashes now stand for
differentiation with respect to the variable~$u$. Note that the above relation was
also obtained by de Souza Dutra~\cite{dutra93} by effecting a transformation $x =
f(u)$ in the Schr\"odinger equation~(\ref{eq:schrodinger}), and redefining the wave
function as $\psi(x) = \sqrt{df(u(x))/du}\, \chi(u)$. However
Eqs.~(\ref{eq:mod-schrodinger}) and~(\ref{eq:E-V}) were first written down by
Bhattacharjee and Sudarshan~\cite{sudarshan} in 1962, and contained, as we have
just shown, de Souza Dutra's result as a particular case ($Q=0$).\par
%
%
In recent papers~\cite{dutra93,dutra94,grosche,dutt}, the
result~(\ref{eq:E_T-V_T}) has been exploited by a number of authors to get
CES~potentials in the form~$V(f(u))$ by choosing judiciously the transformation
function~$f(u)$, and assigning to~$V_T$ an ES~potential with known energy
spectrum and eigenfunctions.\par
%
%
Here we shall exploit it in a slightly different way to get QES~potentials by taking
so(2,1) as the potential algebra of the Schr\"odinger equation, and the resulting
form of the potential~$V_m$, given by (\ref{eq:V_m}), as $V_T$. We get in this way
\begin{equation}
  \left(\case{1}{4} - m^2\right) \left(1 - F^2\right) - 2mFG + G^2 - E_T =
  [f'(u)]^2 [V(f(u)) - E] - \case{1}{2} \Delta V(f(u)), \label{eq:basic-eq}
\end{equation}
where the derivatives of~$F$ and~$G$ have been removed by making use of
Eq.~(\ref{eq:so(2,1)-cond}).\par
%
%
Whereas $F(x)$ and $G(x)$ are known for the three cases summarized
in~(\ref{eq:3cases}), which are relevant for the potential algebra so(2,1), the
mapping function~$f(u)$ is unknown in~(\ref{eq:basic-eq}). In the following, we
propose to use a mapping function that transforms a half-line onto itself. It may be
remarked that in Ref.~\cite{dutt}, a form for $f(u)$ was employed which switched
variable~$x$ to~$u$ transforming in the process the full line to the half-line.\par
%
%
Let us choose the mapping function to be
\begin{equation}
  x = f(u) = \left(e^u - 1\right)^{-1}.   \label{eq:f(u)}
\end{equation}
It is obvious from~(\ref{eq:f(u)}) that $(0, \infty)$ is the domain of both
variables~$x$ and~$u$. Using~(\ref{eq:f(u)}), the Schwartzian
derivative~(\ref{eq:Delta-spec}) reduces to $\Delta V(f(u)) = -1/2$.\par
%
%
Consider first the class~I solutions obtained in~(\ref{eq:3cases}). Substituting
these into~(\ref{eq:basic-eq}), and setting $E=0$ give
\begin{eqnarray}
  V(f(u)) & = & \left(1 - 4 m^2 + 4 b^2\right) \sech u (\sech u - 2) - 8 m b (\tanh u
           \sech u - 2 \tanh u + \sinh u) \nonumber \\
  & & \mbox{} - \left(4 E_T + 1\right) \cosh u (\cosh u - 2)
           - 4 \left(E_T + m^2 - b^2\right).  \label{classI-prel}
\end{eqnarray}
The expression~(\ref{classI-prel}) may be translated in terms of the variable~$x$
through the use of~(\ref{eq:f(u)}), leading to
\begin{equation}
  \mbox{(I)} \qquad V(x) = - \frac{A}{\left(2x^2 + 2x + 1\right)^2} - B \frac{2x + 1}{x
  (x+1) \left(2x^2 + 2x + 1\right)^2} + \frac{C}{x^2 (x+1)^2},   \label{eq:classI} 
\end{equation}
with
\begin{equation}
  A = 4 \left(m^2 - b^2\right) - 1, \qquad B = 4mb, \qquad C = - E_T - \case{1}{4} =
  \left(k - \case{1}{2}\right)^2 - \case{1}{4} = k (k-1).
\end{equation}
\par
%
%
This potential is our first example of QES potential with known $E=0$ eigenvalue.
For nonnegative values of~$B$ and~$C$, its behaviour at the origin is similar to that
of the (shifted) Coulomb effective potential $V_E(x) = \left(\hbar^2/2m\right)
\left(l(l+1)/x^2\right) - \left(e^2/x\right)$. From~(\ref{eq:classI}), we indeed note
that as $x\to 0$, the first term $\to - A$, the second term $\sim - B/x$, and the
third term $\sim C/x^2$. On the other hand, for $x\to \infty$, $V(x)\to 0$.\par
%
%
To consider whether the $E=0$ quantal state is normalizable, let us consider for
instance the potential~(\ref{eq:classI}) for which $m=k$ (i.e., $A$, $B$, and $C$ are
connected by the relation $A = 1 + 4C + 2 \sqrt{1 + 4C} - B^2 \left(1 + \sqrt{1
+ 4C}\right)^{-2}$). The corresponding wave function is $\psi_0(x) = \sqrt{f'(u(x))}\,
\chi_0(u(x))$, where $\chi_0(u)$ is the ground state wave function of class~I
potential in Eq.~(\ref{eq:3cases}), namely $\chi_0(u) \sim G^{k- \frac{1}{2}}\, h$
with $h = \exp\left[b \tanh^{-1}(\sinh u) \right]$~\cite{cq}. Since $\sqrt{f'(u(x))}
\sim \sqrt{2x(x+1)}$, $\psi_0(x) \to 0$ as $x\to 0$, but $\psi_0(x) \sim x$ for
$x\to \infty$. We therefore conclude that for $m=k$, the function~$\psi_0$ is 
non-normalizable.\par
%
%
Turning now to the class~II solutions of (\ref{eq:3cases}), we find
(\ref{eq:basic-eq}) to yield another QES~potential with $E=0$,
\begin{equation}
  \mbox{(II)} \qquad V(x) = \frac{A}{(x+1)^4} - \frac{B}{x (x+1)^3} + \frac{C}{x^2
  (x+1)^2},
\end{equation}
with
\begin{equation}
  A = b^2, \qquad B = 2mb, \qquad C = - E_T - \case{1}{4} = k(k-1).
\end{equation}
\par
%
%
Here, as $x\to \infty$, $V(x) \to 0$, while for $x\to 0$, $V(x) \to A - (B/x) +
(C/x^2)$. Once again $V(x)$ mimics the Coulomb problem near the origin provided
nonnegative values of~$B$ and~$C$ are considered. An analysis of the wave function
corresponding to~$E=0$ for the potential with~$m=k$ (i.e., with
$B A^{-1/2} = 1 + \sqrt{1 + 4C}$) shows that $\psi_0(x) = \sqrt{f'(u(x))}\,
\chi_0(u(x))$, where $\chi_0(u) \sim G^{k- \frac{1}{2}}\, h$, and $h \sim \exp\left(-
b e^{-u}\right)$~\cite{cq}. It is easy to check that $\psi_0 \to \infty$ as $x \to
\infty$, so that in this case too the wave function is non-normalizable.\par
%
%
{}Finally, we take up the class~III solutions of~(\ref{eq:3cases}). For the
corresponding pair $(F, G)$, and $E=0$, $V(x)$ turns out to be
\begin{equation}
  \mbox{(III)} \qquad V(x) = \frac{A}{(2x+1)^2} - \frac{B}{x(x+1)} + \frac{C}{x^2
  (x+1)^2},   \label{eq:classIII}
\end{equation}
with
\begin{equation}
  A = 4 (m+b)^2 - 1, \qquad B = 4mb, \qquad C = - E_T - \case{1}{4} = k(k-1).
  \label{eq:parameters}
\end{equation}
\par
%
%
This QES~potential shares similar qualitative features with the previous ones as
far as its behaviour as $x\to 0$ or $x \to \infty$ is concerned. We shall presently
see that provided certain convergence condition is satisfied, the wave function of
the $E=0$ state turns out to be normalizable for the potentials for which
$m=k$.\par
%
%
To enquire into the normalizability  of the wave function, we note that for the
class~III solutions of~(\ref{eq:3cases}), and $m=k$~\cite{cq},
\begin{equation}
  \psi_0(x) \sim \sqrt{x(x+1)} [\cosech u(x)]^{k - \frac{1}{2}} [
  \tanh(\case{1}{2}u(x)]^b \sim \frac{[x(x+1)]^k}{(2x+1)^{k+b-\frac{1}{2}}}.  
  \label{eq:psi0}
\end{equation}
Since we are dealing with the one-dimensional Schr\"odinger
equation~(\ref{eq:schrodinger}) on the half line, the normalization integral is given
by
\begin{equation}
  \int_0^{\infty} |\psi_0(x)|^2 \, dx \sim \int_0^{\infty} \frac{[x(x+1)]^{2k}}
  {(2x+1)^{2k+2b-1}}\, dx.   \label{eq:normalization}
\end{equation}
To examine its convergence, we make use of the criterion that if $\lim_{x \to
\infty} x^{\alpha} f(x) = L$, where $\alpha>1$, then $\int_a^{\infty} f(x)\, dx$
converges absolutely; if $\lim_{x \to \infty} x^{\alpha} f(x) = L \ne 0$ and $\alpha
\le 1$, then $\int_a^{\infty} f(x)\, dx$ diverges. In the present case, $f(x) \sim
x^{2k-2b+1}$ for $x\to \infty$. So if we take $\alpha = 2b-2k-1$, we have $\lim_{x
\to \infty} x^{\alpha} f(x) = L \ne 0$ and the integral converges if $2b-2k-1 > 1$ or
$b > k+1$, and diverges if $b \le k+1$. Whenever the convergence condition $b > k+1$
is satisfied, the normalization integral~(\ref{eq:normalization}) may be easily
evaluated. For $2k \in \N$, one finds
\begin{equation}
  \int_0^{\infty} |\psi_0(x)|^2 \, dx \sim 2^{-4k-2} \sum_{m=0}^{2k} (-1)^m 
  \left(\begin{array}{c} 
            2k \\ m 
           \end{array}\right) 
  (b + m - k - 1)^{-1}.
\end{equation}
Similar results may be found for other values of~$k$.\par 
%
%
The convergence condition can also be expressed in terms of the parameters
appearing in the potential~(\ref{eq:classIII}). For $m = k + n$,
\begin{eqnarray}
  A & = & (2k + 2b + 2n + 1) (2k + 2b + 2n - 1), \\
  B & = & 4 (k + n) b, \\
  C & = & k (k - 1).   \label{eq:C}
\end{eqnarray}
These imply
\begin{eqnarray}
  k & = & \case{1}{2} \left(1 + \sqrt{1 + 4C}\right),  \label{eq:k} \\
  b & = & \frac{B}{4n + 2 + 2\sqrt{1 + 4C}}, \\
  A & = & \left(2k + 2n + 1 + \frac{B}{2k+2n}\right) \left(2k + 2n - 1 +
         \frac{B}{2k+2n}\right) \nonumber \\   
  & = & \left(2n + 2 + \sqrt{1 + 4C} + \frac{B}{2n + 1 + \sqrt{1 + 4C}}\right)
         \nonumber \\
  & & \mbox{} \times \left(2n + \sqrt{1 + 4C} + \frac{B}{2n + 1 + \sqrt{1 +
         4C}}\right).     \label{eq:A} 
\end{eqnarray}
In Eq.~(\ref{eq:k}), we selected the solution of Eq.~(\ref{eq:C}) satisfying, for
nonnegative values of~$C$, the requirement imposed on so(2,1) unitary irreducible
representation labels~$k$, namely $k>0$. We note that we actually obtain $k\ge 1$.
For the case $m=k$ or $n=0$, the convergence condition $b > k+1$ then reads 
\begin{equation}
  B > 4k (k+1) = 4 \left(1 + C + \sqrt{1 + 4C}\right),   \label{eq:convergence}
\end{equation}
which in turn places non-trivial restriction on the run of $A$, as we shall now
proceed to show.\par
%
%
The zeroes, as well as the regions of positivity and negativity, of 
potentials~(\ref{eq:classIII}) are displayed in Table~1 in terms of $A$, $B$,
and~$C$, for $-\infty < A < \infty$, $0\le B < \infty$, and $0\le C < \infty$. Various
types of potentials may be distinguished according to their behaviour over $(0,
\infty)$, and/or the parameter range.\par
%
%
By using Eqs.~(\ref{eq:C}), and~(\ref{eq:A}) (for $n=0$), expressing $C$ and~$A$ in
terms of~$k$ and~$B$, it is now straightforward to show that those
potentials~(\ref{eq:classIII}) satisfying the convergence
condition~(\ref{eq:convergence}) belong to either type~I or type~V of Table~1,
according to whether $k>1$ or $k=1$ (equivalently $C>0$ or
$C=0$). Examples for both types are displayed in Figs.~1 and~2,
respectively.\par
%
%
It appears that contrary to the effective Coulomb potential, type~I and~V
potentials approach zero from the top as $x\to \infty$. This explains why the
corresponding $E=0$ eigenstates are normalizable (and bound). Type~I potentials
behave as the effective Morse or spherical well potentials, which may provide
phenomenological descriptions of alpha-decay, while type~V potentials have the
features of one of the effective radial power law potentials with $E=0$
normalizable states found by Daboul and Nieto~\cite{daboul}.\par
%
%
{}For simplicity's sake, the discussion of the $E=0$ wave function normalizability
has been carried out for those potentials for which $m=k$ or $n=0$. It is clear
however that its feasibility is by no way restrited to such values. For the class~III
potentials~(\ref{eq:classIII}), for instance, the counterpart of Eq.~(\ref{eq:psi0})
for $m=k+n$, and any $n \in \N$ can be found from the singular Gendenshtein
potential wave functions given in Ref.~\cite{cooper}. It is given by
\begin{equation}
  \psi_n(x) \sim \frac{[x(x+1)]^m}{(2x+1)^{m+b-\frac{1}{2}}} P_n^{(b-m,-b-m)}
  \left(\frac{2x^2+2x+1}{2x(x+1)}\right),
\end{equation}
where $m$, $b$, and $n$ are related to $A$, $B$, and~$C$
through~(\ref{eq:parameters}), and $P_n^{(b-m,-b-m)}$ denotes a Jacobi
polynomial.\par
%
%
In conclusion, we did provide some new examples of QES~potentials with known
$E=0$ quantal states, and we did show that some of the latter may be normalizable
and bound. It should be remarked that contrary to Daboul and Nieto, no example of
normalizable, but unbound states has been found.\par
%
%
We constructed our QES~potentials by a method inspired from general techniques for
generating ES and CES~potentials, and we used as starting point some ES~potentials
for which the Schr\" odinger equation admits an so(2,1) potential algebra. The
method proposed here might also be useful for generating QES~potentials from
some other ES~potentials.\par
%
%
One of us (B.B.) thanks Professor C. Quesne for warm hospitality at Universit\'e
Libre de Bruxelles, where this work was done. B.B.'s work was supported in part by
the Council of Scientific and Industrial Research, New Delhi, through the grant of a
project. He also acknowledges travel support received from the University of
Calcutta and Sir Dorabji Tata Trust, Bombay.\par
\newpage
%
%

\newpage
%
%
\begin{table}[h]

\caption[table]{Zeroes and sign of the potential $V(x)$, defined in
Eq.~(\ref{eq:classIII}), for $-\infty < A < \infty$, $0 \le B < \infty$, and $0 \le C <
\infty$. The quantities $x_{0+}$ and $x_{0-}$ are defined by $ x_{0\pm} \equiv
\frac{1}{2}\left(-1 + \sqrt{X_{\pm}/Y}\right)$, where $X_{\pm} \equiv A - 2B
- 8C \pm 2\sqrt{\Delta}$, $Y \equiv A - 4B$, and $\Delta \equiv (B + 4C)^2 - 4AC$.}

\vspace{1cm}
\begin{tabular}{llll}          
  \hline\\[-0.2cm] 
  Type & Parameters & Zeroes & Sign
  \rule[-0.3cm]{0cm}{0.6cm}\\[0.3cm]
  \hline\\[-0.2cm] 
  I & $0 \ne 16C < 4B < A < \frac{(B + 4C)^2}{4C}$ & $x_{0+}$, $x_{0-}$ & $>0$ if $x <
        x_{0-}$ or $x > x_{0+}$ \\[0.5cm]
  & & & $<0$ if $x_{0-} < x < x_{0+}$ \\[0.5cm] 
  II & $0 \ne 16C < A = 4B$ & $x_0 = - \frac{1}{2} + \frac{1}{2}
        \sqrt{\frac{B}{B-4C}}$ & $>0$ if $x < x_0$ \\[0.5cm]
  & & & $<0$ if $x > x_0$ \\[0.5cm]
  III & $C\ne 0$, $A < \min\left(4B, \frac{(B+4C)^2}{4C}\right)$ & $x_{0-}$ & $>0$ if
        $x < x_{0-}$ \\[0.5cm]
  & & & $<0$ if $x > x_{0-}$ \\[0.5cm]
  IV & $0 \ne 4C < B$, $A = \frac{(B+4C)^2}{4C}$  & $x_0 = - \frac{1}{2} + \frac{1}{2}
        \sqrt{\frac{B+4C}{B-4C}}$ & $>0$ if $x \ne x_0$ \\[0.5cm]
  V & $C = 0$, $0 \ne 4B < A$ & $x_0 = - \frac{1}{2} + \frac{1}{2}
        \sqrt{\frac{A}{A-4B}}$ & $<0$ if $x < x_0$ \\[0.5cm]
  & & & $>0$ if $x > x_0$ \\[0.5cm] 
  VI & $A = 4B$, $0 \ne 4C \ge B$ & \trait & $>0$ \\[0.5cm]
  VII & $C \ne 0$, $4B \ne A > \frac{(B+4C)^2}{4C}$ & \trait & $>0$ \\[0.5cm] 
  VIII & $0 \ne 4C > B$, $A = \frac{(B+4C)^2}{4C}$ & \trait & $>0$ \\[0.5cm]
  IX & $B < 4C \ne 0$, $4B < A <  \frac{(B+4C)^2}{4C}$ & \trait & $>0$ \\[0.5cm]
  X & $C = 0$, $A \le 4B \ne 0$ & \trait & $<0$ \\[0.5cm] 
  XI & $B = C = 0$, $A > 0$ & \trait & $> 0$, $V(0) < \infty$ \\[0.5cm]
  XII & $B = C = 0$, $A < 0$ & \trait & $< 0$, $V(0) > -\infty$ \\[0.5cm]
  \hline 
\end{tabular}
\end{table}
\newpage
%
%
\section*{Figure captions}
\begin{figure}[h]
\caption{The potential $V(x)$ of Eq.~(\ref{eq:classIII}) for $A=399$, $B=64$, $C=2$,
or $k=m=2$, $b=8$.}
\end{figure}
\begin{figure}[h]
\caption{The potential $V(x)$ of Eq.~(\ref{eq:classIII}) for $A=63$, $B=12$, $C=0$,
or $k=m=1$, $b=3$.}
\end{figure}

\end{document}